# Face masks, vaccination rates and low crowding drive the demand for the London Underground during the COVID-19 pandemic

Prateek Bansal[1*], Roselinde Kessels[2], Rico Krueger[3], Daniel J Graham[1]

[1]Transport Strategy Centre, Imperial College London, UK
[2]Department of Data Analytics and Digitalization, Maastricht University, the Netherlands
[3]Transport and Mobility Laboratory, École Polytechnique Fédérale de Lausanne, Switzerland

*corresponding author (Email: pb422@cornell.edu)

## Abstract

The COVID-19 pandemic has drastically impacted people's travel behaviour and out-of-home activity participation. While countermeasures are being eased with increasing vaccination rates, the demand for public transport remains uncertain. To investigate user preferences to travel by London Underground during the pandemic, we conducted a stated choice experiment among its pre-pandemic users (N=961). We analysed the collected data using multinomial and mixed logit models. Our analysis provides insights into the sensitivity of the demand for the London Underground with respect to travel attributes (crowding density and travel time), the epidemic situation (confirmed new COVID-19 cases), and interventions (vaccination rates and mandatory face masks). Mandatory face masks and higher vaccination rates are the top two drivers of travel demand for the London Underground during COVID-19. The positive impact of vaccination rates on the Underground demand increases with crowding density, and the positive effect of mandatory face masks decreases with travel time. Mixed logit reveals substantial preference heterogeneity. For instance, while the average effect of mandatory face masks is positive, preferences of around 20% of the pre-pandemic users to travel by the Underground are negatively affected. The estimated demand sensitivities are relevant for supply-demand management in transit systems and the calibration of advanced epidemiological models.

# 1 Introduction

The COVID-19 pandemic has severely affected the demand for public transport in many parts of the world (Tirachini & Cats 2020; Transport Strategy Centre 2020; Vickerman 2021). Initial lockdown measures resulted in a sharp drop in ridership (Transport Strategy Centre 2020). Demand for public transport has increased since the early stages of the pandemic but remains below pre-COVID levels, possibly due to decreased out-of-home activity participation and a perceived risk of infection in enclosed public spaces (Tirachini & Cats 2020; Transport Strategy Centre 2020; Vickerman 2021). Despite increasing vaccination rates and the gradual easing of restrictions, the trajectory of future public transport demand is uncertain.

SARS-CoV-2, the virus that causes COVID-19, spreads via droplets and aerosols (Anderson et al. 2020; Asadi et al. 2020; Morawska & Cao 2020). As a consequence, COVID-19 transmissions are more likely to occur in enclosed spaces such as public transport (Morawska & Milton 2020; Prather et al. 2020). To counteract the spread of COVID-19, authorities have mandated various non-pharmaceutical interventions, such as closures of schools and non-essential businesses, work-from-home, and restrictions to out-of-home activities. Interventions such as physical distancing requirements, mandatory face masks and stricter hygiene protocols aim at reducing the risk of contagion in public spaces, including public transport. Besides, information campaigns of public health authorities have emphasised that such measures can effectively prevent COVID-19 transmissions. But, at the same time, such measures have affected the demand for public transport. In some jurisdictions, authorities even discouraged the use of public transport (see Tirachini & Cats 2020, and the literature referenced therein).

To this date, there is no conclusive evidence about the likelihood of COVID-19 infections on public transport compared to other places (Hörcher et al. 2021). However, it is well established that face masks effectively reduce the risk of contagion when adequate physical distancing cannot be maintained (Cheng et al. 2021). In addition, COVID-19 vaccines provide effective protection against symptomatic and asymptomatic infections with the virus (Dagan et al. 2021; Hall et al. 2021). Regardless of the objective effectiveness of these measures, travellers may still perceive that travelling by public transport presents a significant infection risk, especially in crowded conditions. Wearing a mask in long-distance travel might also discourage the use of public transport. Thus, perceived infection risks and the absence or presence of countermeasures are likely to affect the demand for public transport, both during and after the COVID-19 pandemic.

Knowledge of public transport users' sensitivities to crowding, the epidemic situation and non-pharmaceutical interventions is essential both in the short run and in the long run in a post-pandemic world. First, insights into demand characteristics may inform the design and implementation of demand management strategies aimed at reducing crowding levels in public transport (Hörcher et al. 2021; Tirachini & Cats 2020). Second, advanced person-centric epidemic modelling and simulation frameworks (e.g. Aleta et al. 2020; Müller et al. 2021), which account for time- and place-specific infection risks as well as various aspects of activity-travel behaviour, depend on detailed information about demand characteristics in order to accurately forecast the progression of the epidemic situation. Finally, insights into demand characteristics are crucial for supply-side planning and management, project appraisal and welfare analysis.

A substantial body of literature investigated crowding sensitivities of subway demand prior to the COVID-19 pandemic (see, Bansal et al. 2019, 2020; Li & Hensher 2011; Tirachini et al. 2017; Wardman & Whelan 2011, for reviews). Crowding is typically measured in terms of passenger densities, vehicle occupancy rates and the availability of empty seats or standing room. In general, higher levels of crowding are found to increase the disutility of travel time. The effect of crowding on travel time disutility can be quantified using crowding multipliers which capture the trade-offs between crowding and travel time. In a recent international comparison of crowding multipliers, Tirachini et al. (2017) find that at a crowding density of four standing passengers per square meter, travel time is perceived between 1.15 and 2 times more costly than in a baseline scenario with zero standing passengers per square meter. To the best of our knowledge, no study to this date has investigated crowding sensitivities in the context of a pandemic and the COVID-19 pandemic, specifically.

To address this gap in the literature, we conducted a stated preference experiment among the pre-COVID users of the London Underground and analysed the collected data using multinomial and mixed logit models. Our analysis provides insights into the sensitivity of the demand for the London Underground with respect to travel attributes (crowding density, travel time), the epidemic situation (confirmed new COVID-19 cases), non-pharmaceutical interventions (whether masks are mandatory or not), and pharmaceutical measures (vaccination rates). To illustrate the practical significance of the considered attributes, we analyse their effects on the probability of using the Underground by its pre-pandemic users. Variations in these probabilities across users with different demographic characteristics and opinions about countermeasures are also presented.

We organise the remainder of this paper as follows. In Section 2, we describe the methodological approach of the study. In Section 3, we present the estimation results and analyse the practical significance of attributes and preference heterogeneity. Finally, we conclude in Section 4.

## 2 Methodology

### 2.1 Survey and experimental design

We conducted a stated preference survey among pre-pandemic users of the London Underground between March and May 2021. The survey included questions regarding respondents' current travel behaviour, exposure to COVID-19, vaccination status, socio-demographic characteristics as well as regarding perceptions about COVID-19 vaccination, face masks and the UK government's approach to handle the pandemic. The main part of the survey consisted of a discrete choice experiment (DCE) to understand the preferences of users of the London Underground during the pandemic. In the DCE, the respondents were provided with choice scenarios based on crowding levels in the London Underground, travel time, daily new confirmed COVID-19 cases (7-day rolling average), the COVID-19 vaccinated population and mask norms. In each scenario, the respondents were asked to choose their preferred travel profile among two options involving travel by London Underground. An example of a choice scenario is presented in Figure 1.

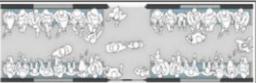

| Attributes | Travel profile 1 | Travel profile 2 |
|---|---|---|
| Crowding levels | 1 person / square metre | 6 persons / square metre |
| Standing in the tube? | No | No |
| In-vehicle (on-board) travel time **(minutes)** | 21 minutes | 30 minutes |
| Daily new COVID-19 cases in the UK **(per 100 thousand inhabitants)** | 50 | 10 |
| Mask compulsory? | Yes | Yes |
| Vaccinated population in the UK | 50% | 80% |

**Figure 1.** An example of a choice scenario presented to respondents.

For the DCE, we adopted a partial profile design with three blocks and eight choice situations per block (Kessels et al., 2011; Kessels et al., 2015). One of three blocks was randomly chosen and presented to a respondent. All blocks were distributed uniformly over the respondents. The levels of the considered attributes in the DCE are presented in Table 1. The maximum crowding density level is six persons per square meter, which is the technical capacity of the London Underground. Travel time is pivoted on a respondent's travel time in the most frequent trip by the Underground. The highest level of daily new confirmed cases (7-day rolling average) is 90 per 100,000 inhabitants, which is the maximum observed number of daily new cases in the UK, which occurred in January 2021. We chose the remaining attribute levels based on design judgement. Technical details of the partial profile design are presented in the Appendix.

**Table 1.** Attribute levels of the discrete choice experiment design.

| Attributes | Travel profile 1 | Travel profile 2 |
|---|---|---|
| Crowding density (persons per square meter) | [0,1,2,4,6] | [0,1,2,4,6] |
| Standing in the Underground? | [Yes, No] | [Yes, No] |
| Travel time (minutes) | [-30%, current,15%,30%] | [-30%, current,15%,30%] |
| Daily new COVID-19 cases (per $10^5$) | [10,30,50,70,90] | [10,30,50,70,90] |
| Mask compulsory? | [Yes, No] | [Yes, No] |
| Vaccine adoption in the UK | [5%,20%,35%,50%,65%,80%] | [5%,20%,35%,50%,65%,80%] |

## 2.2 Data collection

To ensure that the respondents were familiar with the London Underground and hypothetical bias in the DCE remained minimal, we set strict eligibility criteria for the survey participation. Londoners older than 18 years, who had used the London Underground for three or more round trips per week in 2019, had spent more than nine months in London during 2020, and intended to stay more than nine months in London during 2021 were eligible for the survey. The study was reviewed and approved by the Research Governance and Integrity Team at Imperial

College London on March 11, 2021 (SETREC number: 21IC6629) under the Science Engineering Technology Research Ethics Committee (SETREC) process.

In total, 1080 responses were collected. To maintain data quality, we performed several checks. After excluding fast responses (i.e., with a response time below forty percent of the median response time), straight-liners (i.e., chose the same travel profile number across all eight choice situations in the DCE), and inconsistent responses (i.e., reported household size lower than the sum of the number of children and workers), 961 responses remained for further analysis. Table 2 shows that the sample is representative of the London population across gender and age groups.

**Table 2.** London population and sample proportions across different demographic groups.

|  | 2011 United Kingdom census (London population) | Sample proportions (N=961) |
|---|---|---|
| **Gender** | | |
| Male | 49.5% | 48.3% (464) |
| Female | 50.5% | 51.7% (497) |
| **Age** | | |
| 19 to 39 years | 47.4% | 41.7% (401) |
| 40 to 59 years | 32.0% | 36.3% (349) |
| 60+ years | 21.7% | 22.0% (211) |

## 2.3 Summary statistics

Travel mode preferences of commuters before and during the pandemic are presented in Table 3. The table shows that around 45% of workers did not commute during the pandemic, whereas this proportion was below 2% before the pandemic. Among workers who commuted, the share of the London Underground has dropped by around 25-30% during the pandemic. At the same time, shares of sustainable travel modes and cars have increased by 10-15%. These statistics suggest that understanding the factors that can help regain the demand for the London Underground is the need of this hour.

**Table 3.** Travel mode preferences of workers before and during the COVID-19 pandemic (N=798).

| Travel mode | All workers | | | Workers who commuted | | |
|---|---|---|---|---|---|---|
|  | Nov-19 | Jun-20 | Jan-21 | Nov-19 | Jun-20 | Jan-21 |
| Bus/tram | 74 | 54 | 62 | 9% | 13% | 14% |
| Underground | 542 | 168 | 190 | **69%** | **39%** | **42%** |
| Car | 39 | 69 | 67 | **5%** | **16%** | **15%** |
| Taxi/ridesharing/carpool | 4 | 14 | 16 | 1% | 3% | 4% |
| Bicycle/walk | 32 | 84 | 73 | **4%** | **20%** | **16%** |
| Rail | 93 | 38 | 41 | 12% | 9% | 9% |
| I did not travel to work | **14** | **371** | **349** | | | |

Table 4 summarises relevant socio-demographic and COVID-related characteristics. The results show that only 52% of respondents are satisfied with the UK government's approach to handle the pandemic, 9% of respondents tested COVID positive, and 55% of respondents have taken at least one dose of the vaccine.

Table 4. Sample summary statistics (N=961).

| Variables | Sample proportions |
|---|---|
| Monthly household income above 10,000 pounds? | 26.8% |
| Full time employed? | 54.2% |
| Unemployed or retired? | 16.9% |
| Married? | 37.5% |
| Asian or Asian British? | 14.7% |
| Have at least a bachelor's degree? | 66.2% |
| Satisfied with government's approach to COVID-19? | 52.1% |
| Tested COVID positive? | 9.0% |
| Took at least one dose of a vaccine? | 55.5% |

Table 5 presents a summary of attitudes regarding COVID-19 vaccines and masks. 87% and 81% of respondents trust vaccine technology and the information provided by the UK government regarding vaccines, respectively. However, 51-53% of respondents are worried about the vaccines' side effects and do not find vaccines powerful in the long term. Despite that 38% of respondents believe in possibilities of respiratory issues due to mask usage, 94% of respondents wear masks in public spaces. 79% of respondents believe in the abilities of a mask to control the COVID-19 spread.

Table 5. Attitudes regarding COVID-19 vaccines and masks.

| | Agree | Somewhat agree | Somewhat disagree | Disagree |
|---|---|---|---|---|
| **Vaccine-related statements** | | | | |
| I trust COVID-19 vaccine technology. | **53%** | **34%** | 7% | 5% |
| I am worried about the side-effects of the COVID-19 vaccine. | 22% | 29% | 27% | 22% |
| I think the vaccine will not be powerful in the long term. | 16% | 37% | 33% | 13% |
| The information provided by the government regarding COVID-19 vaccines is credible. | **33%** | **48%** | 14% | 5% |
| **Mask-related statements** | | | | |
| Any kind of mask helps in controlling COVID-19 spread. | **40%** | **39%** | 14% | 8% |
| I always wear a mask at public spaces. | **77%** | **17%** | 4% | 3% |
| Wearing a mask can cause respiratory issues. | 11% | 27% | 33% | 30% |

We also explore heterogeneity in respondents' opinions across their vaccination status in Table 6. The results indicate that only 2-4% of those respondents who have taken at least one dose of a vaccine do not trust vaccine technology, but this proportion is much higher (24%) for unvaccinated respondents. This statistic shows that around one-fourth of unvaccinated respondents might be reluctant to be vaccinated. Similarly, the proportion of respondents satisfied with the UK government's approach towards the COVID-19 pandemic increases with the vaccination status. More specifically, while only 45% of the unvaccinated respondents are satisfied with the government's approach, 57% of the respondents who have taken the first dose of a vaccine and 63% of the fully vaccinated respondents are satisfied.

**Table 6.** Trust in COVID-19 vaccine technology and satisfaction with government's approach among respondents with different vaccination levels.

|  | First dose (N=435) | Both doses (N=98) | No dose (N=428) |
|---|---|---|---|
| *I trust COVID-19 vaccine technology* | | | |
| Agree | 96% | 98% | **76%** |
| Disagree | 4% | 2% | **24%** |
| *Satisfaction with government's approach to the COVID-19 pandemic* | | | |
| Satisfied | 57% | 63% | **45%** |
| Dissatisfied | 43% | 37% | **55%** |

## 2.4 Modelling approach

The multinomial logit (MNL) model is the base model in analysis of the stated choice data. The model defines the utility that respondent $n$ ($n = 1, …, N$) attaches to travel profile $j$ ($j = 1, 2$) in choice situation $s$ ($s = 1, …, 8$) as the sum of a systematic and a stochastic component:

$$U_{njs} = V_{njs} + \varepsilon_{njs}. \tag{1}$$

The stochastic component $\varepsilon_{njs}$ is the idiosyncratic error term, which is assumed to be independent and identically distributed across $n$, $j$, $s$ according to a standard Gumbel distribution.

In our analysis, we adopt different specifications of the systematic component of the utility. We first investigate the main effects of the six alternative-specific attributes plus an alternative-specific constant (*ASC*):

$$V_{njs} = ASC + \sum_{k=1}^{6} \beta_k x_{njsk} = ASC + \boldsymbol{\beta}' \boldsymbol{x}_{njs}, \tag{2}$$

where $\boldsymbol{x}_{njs}$ is a 6-dimensional vector containing the attribute levels $x_{njsk}$ of travel profile $j$ in choice situation $s$ for respondent $n$ and $\boldsymbol{\beta}$ is a 6-dimensional vector of parameter values $\beta_k$ representing marginal utilities of the attributes.

In another utility specification, we also add all possible two-way interactions between the alternative-specific attributes. For both utility specifications, the MNL probability that respondent $n$ chooses travel profile $j$ in choice situation $s$ is obtained using the following closed-form expression:

$$p_{njs} = \frac{\exp(V_{njs})}{\exp(V_{n1s}) + \exp(V_{n2s})}. \tag{3}$$

To explore the unobserved preference heterogeneity in the main effects, the more flexible panel mixed (MXL) model is adopted. Inclusion of possible two-way interactions between the alternative-specific attributes leads to the following systematic utility specification of MXL:

$$V_{njs} = ASC + \boldsymbol{\beta}'_n \boldsymbol{x}_{njs} + \sum_{\forall k} \sum_{\forall l, l \neq k} \delta_{kl} x_{njsk} x_{njsl}, \tag{4}$$

where $\boldsymbol{\beta}_n$ is randomly drawn from a mixing distribution $f(\boldsymbol{\beta}_n)$ (e.g., normal and lognormal distributions), and $\delta_{kl}$ is the marginal utility of the interaction between attributes *k* and *l*. In another utility specification of MXL, interactions between alternative-specific attributes and individual-level covariates (i.e., demographic characteristics and opinions) are included to capture observed heterogeneity in the mean effects of the attributes. The MXL probability that respondent *n* chooses alternative *j* in choice situation *s* is obtained by solving the following 6-dimensional integral:

$$p_{njs} = \int \frac{\exp(V_{njs})}{\exp(V_{n1s}) + \exp(V_{n2s})} f(\boldsymbol{\beta}_n) d\boldsymbol{\beta}_n. \tag{5}$$

The probability expression in Eq. 3 or Eq. 5 is used to compute the loglikelihood, which is maximised to obtain estimates of the model parameters. We estimate all models using the gmnl package in R (Sarrias and Daziano, 2017).

## 3 Results

In this section, we present the estimation results of the MNL and MXL models. The parameter estimates provide insights into the statistical significance of the predictors and the extent of preference heterogeneity. We also illustrate the practical significance of the predictors by plotting their effect on the probability to use the Underground during the pandemic.

### 3.1 Multinomial logit

We consider two specifications of the MNL model, the results of which appear in Table 7. Whereas specification 1 evaluates the main effects of alternative-specific attributes, specification 2 explores interaction effects along with main effects. The results of specification 1 reveal that the main effect of each attribute is statistically significant at a 0.05 level and has the expected sign. Whereas crowding density, standing in the Underground, travel time, and daily new COVID-19 cases negatively affect the respondents' likelihood to choose the Underground, the effects of mandatory face masks and higher rates of vaccination are positive. To obtain an initial understanding of the explanatory power of each attribute, we estimate multiple models while keeping one attribute at a time along with the alternative-specific constant. The increase in the model fit over the constant-only model shows the empirical relevance of the attribute. The results indicate that mandatory face masks and higher rates of vaccination are the top two drivers in loglikelihood improvement (from -5318.0 to -4871.2 and -5195.4), whereas the addition of the standing attribute leads to virtually no gain in the loglikelihood (from -5318.0 to -5317.7). The likelihood ratio test suggests that the constant-only model is superior to the one with the constant and standing attribute (Chi-square statistic: 0.49, Df: 1, p-value: 0.48).

In specification 2, we examine various utility structures with possible two-way interactions between attributes. Only three interaction effects are significant at a 0.1 level, while the main effect of the standing attribute loses statistical significance after including these interactions[1]. The likelihood ratio test suggests that specification 2 is statistically superior to specification 1 in terms of goodness of fit (Chi-square statistic: 39.1, Df: 2, p-value: 3.3e-09). We observe that the positive effect of mandatory face masks on the propensity to choose the London Underground decreases in travel time, most probably because the discomfort due to wearing a mask increases with the travel duration. Moreover, the positive effect of vaccination rates increases with crowding density. This result is aligned with the intuition that the risk of COVID-19 transmission increases with crowding levels, and that countermeasures become critical in such situations. Note that the negative impact of new COVID-19 cases increases with travel time as the risk of being infected with COVID-19 increases with travel duration. However, this effect is statistically significant at a 0.07 level and completely loses significance after accounting for the unobserved preference heterogeneity in the MXL model.

**Table 7.** Parameter estimates of two MNL model specifications.

| Explanatory variables | Specification 1 | | | Specification 2 | |
|---|---|---|---|---|---|
| | Estimate | z-value | Loglikelihood | Estimate | z-value |
| Alternative-specific constant (alternative 2) | -0.11 | -3.7 | -5318.0 | -0.15 | -4.9 |
| **Alternative-specific variables (main effects)** | | | | | |
| Crowding density (persons per square meter) | -0.17 | -12.0 | -5255.2 | -0.30 | -8.1 |
| Standing in the Underground? | -0.10 | -2.4 | -5317.7 | | |
| Travel time (minutes/100) | -1.55 | -5.7 | -5291.5 | -0.76 | -2.6 |
| Daily new COVID cases (per $10^7$) | -1.08 | -9.1 | -5307.5 | -0.58 | -3.4 |
| Mask compulsory? | 0.74 | 19.0 | -4871.2 | 1.01 | 16.8 |
| Vaccine adoption (%) | 1.51 | 16.5 | -5195.4 | 1.09 | 9.5 |
| **Interaction between alternative-specific variables** | | | | | |
| Travel time x Mask compulsory? | | | | -0.63 | -4.9 |
| Crowding density x Vaccine adoption (%) | | | | 0.26 | 4.2 |
| Travel time x Daily new COVID cases | | | | -0.62 | -1.8 |
| Loglikelihood | | | -4664.9 | | -4645.4 |

**Note:** The "Loglikelihood" column presents loglikelihood values for a model with one alternative-specific variable and the constant. A comparison of these values shows the practical relevance of the variables in explaining respondent preferences.

### 3.2 Mixed logit

We consider two specifications of the MXL model whose parameter estimates are presented in Table 8. Specification 1 estimates the main and interaction effects of alternative-specific attributes and incorporates unobserved taste heterogeneity in the main effects using parametric mixing distributions. On top of specification 1, specification 2 also includes the heterogeneity in the mean attribute effects across different groups segmented based on respondents' socio-demographics and opinions. Whereas specification 1 is more suitable for forecasting because it does not involve opinion-based covariates, specification 2 is beneficial from an exploratory perspective as it offers more insights into preference heterogeneity. The likelihood ratio test suggests that specification 2 of the MXL model is superior to specification 1 in terms of goodness of fit (Chi-square statistic: 280.5, Df: 20, p-value < 2.2e-16), which is superior to specification 2 of the MNL model (Chi-square statistic: 498.6, Df: 4, p-value < 2.2e-16).

---

[1] Previous studies related to crowding valuation of metro/subway passengers have found statistically significant interactions between crowding, travel time, and standing attributes (see, e.g., Bansal et al. 2019). These studies use interaction effects to derive crowding and standing multipliers. However, we do not find these interaction effects to be significant in our study, most probably because COVID-19 related countermeasures such as mandatory face masks and rates of vaccination are the main drivers of travel demand for the Underground as compared to crowding levels and travel time (as also illustrated in the "Loglikelihood" column of Table 7).

The marginal utility of mandatory face masks is assumed to follow a normal distribution, because having to wear face masks in the Underground is likely to encourage the use of the Underground, but a segment of respondents might also be negatively affected. Since the increase in crowding density, travel time, and new COVID-19 cases is likely to negatively affect the preferences to travel by the Underground, a lognormal mixing distribution is assumed for the *negative* marginal utilities of these attributes. Similarly, the marginal utility of vaccination rates is considered to follow a lognormal distribution because it is likely to increase the propensity to use the Underground. Note that parameters of the normal distribution associated with the lognormally distributed marginal utilities are obtained as a by-product of the estimation, but they are not directly interpretable. We use parameter estimates of the underlying normal distribution to compute the mean, standard deviation, and various percentiles of the lognormal distribution. The standard errors of the distributional summary statistics are obtained by simulating the asymptotic distribution of the estimated parameters. Distributional summaries and standard errors for both MXL specifications are presented in Table 9.

The results indicate the presence of statistically significant unobserved and observed taste heterogeneity in all attributes. Since the effect of mandatory face masks varies with travel time in MXL's specification 1, the proportion of respondents with negative marginal utility of mandatory face masks is pivoted around travel time. Among users with an Underground travel time of 30 and 90 minutes, the propensity of 17.5% and 26.8% users to travel by the Underground is negatively impacted due to mandatory face masks. The results of MXL's specification 2 indicate substantial heterogeneity in the mean effects of mandatory face masks, vaccination rates, and crowding density across segments of pre-pandemic users with differing demographics and opinions about countermeasures. We evaluate the extent of these effects in the next section.

Table 8. Parameter estimates of two MXL model specifications.

| Explanatory variables | Specification 1 | | Specification 2 | |
|---|---|---|---|---|
| | Estimate | z-value | Estimate | z-value |
| Alternative specific constant (alternative 2) | -0.17 | -4.7 | -0.17 | -4.4 |
| **Random parameters** | | | | |
| *(-) Crowding density (persons per square meter) [Lognormal]* | | | | |
| Mean | -1.06 | -8.7 | -0.76 | -5.3 |
| Std. dev. | 0.87 | 13.9 | 0.67 | 7.2 |
| *(-) Travel time (minutes/100) [Lognormal]* | | | | |
| Mean | -1.32 | -22.1 | 0.63 | 2.2 |
| Std. dev. | 2.41 | 112 | 1.34 | 9.4 |
| *(-) Daily new COVID cases (per $10^7$) [Lognormal]* | | | | |
| Mean | -0.09 | -0.6 | -1.35 | -1.3 |
| Std. dev. | 0.82 | 6.1 | 1.58 | 3.3 |
| *Mask compulsory? [Normal]* | | | | |
| Mean | 1.45 | 20.3 | 0.34 | 1.8 |
| Std. dev. | 1.33 | 21.8 | 1.08 | 13.2 |
| *Vaccine adoption (%) [Lognormal]* | | | | |
| Mean | -0.11 | -1.0 | -0.69 | -3.9 |
| Std. dev. | 1.47 | 25.8 | 1.87 | 29 |
| **Interaction between alternative-specific variables** | | | | |
| (-) Travel time x Mask compulsory? | 0.70 | 5.2 | 0.31 | 1.8 |
| (-) Crowding density x Vaccine adoption (%) | -0.35 | -5.6 | -0.34 | -4.4 |
| **Interaction between alternative- & individual-specific variables** | | | | |
| (-) Crowding density x Age below 40 years? | | | -0.13 | -3.9 |
| (-) Crowding density x Unemployed or retired? | | | -0.11 | -2.6 |
| (-) Crowding density x Satisfied with govt's approach towards COVID? | | | -0.07 | -2.3 |
| (-) Crowding density x Tested positive for COVID? | | | -0.12 | -2.0 |
| (-) Crowding density x Have at least a bachelor's degree? | | | 0.10 | 3.1 |
| (-) Crowding density x Married? | | | -0.11 | -3.3 |
| (-) Travel time x Mask helps in controlling COVID spread? | | | -2.16 | -3.5 |
| (-) Daily new COVID cases x Full time employed? | | | -0.41 | -1.8 |
| (-) Daily new COVID cases x Always wear a mask at public spaces? | | | 0.85 | 2.5 |
| Mask compulsory? x Age below 40 years? | | | -0.55 | -6.0 |
| Mask compulsory? x Male? | | | -0.31 | -3.7 |
| Mask compulsory? x Unemployed or retired? | | | 0.38 | 3.2 |
| Mask compulsory? x Mask helps in controlling COVID spread? | | | 0.49 | 4.8 |
| Mask compulsory? x Always wear a mask at public spaces? | | | 1.33 | 8.1 |
| Mask compulsory? x Wearing a mask can cause respiratory issues? | | | -0.79 | -9.2 |
| Mask compulsory? x Monthly household income > 10,000 pounds? | | | -0.19 | -2.1 |
| Vaccine adoption (%) x Worried about side-effects of COVID vaccine? | | | -0.91 | -4.8 |
| Vaccine adoption (%) x Vaccines won't be powerful in long term? | | | -0.65 | -3.3 |
| Vaccine adoption (%) x Mask helps in controlling COVID spread? | | | 1.40 | 7.4 |
| Vaccine adoption (%) x Asian or Asian British? | | | 0.93 | 3.6 |
| Null (constant-only) Loglikelihood | -5318.0 | | -5318.0 | |
| Loglikelihood | -4397.6 | | -4256.9 | |

**Note:** For lognormally distributed parameters, mean and std. dev. of the underlying normal distribution are presented.

Table 9. Summary of the mixing distributions of the random parameters in two MXL model specifications.

| Random parameters | Specification 1 | | Specification 2 | |
|---|---|---|---|---|
| | Estimate | z-value | Estimate | z-value |
| *(-) Crowding density (persons per square meter) [Lognormal]* | | | | |
| Mean | 0.50 | 13.3 | 0.59 | 9.7 |
| Std. dev. | 0.53 | 15.2 | 0.44 | 6.6 |
| 5th percentile | 0.08 | 4.5 | 0.16 | 3.5 |
| 50th percentile | 0.35 | 8.2 | 0.47 | 6.9 |
| 95th percentile | 1.44 | 20.5 | 1.41 | 10.4 |
| *(-) Travel time (minutes/100) [Lognormal]* | | | | |
| Mean | 4.92 | 103 | 4.68 | 8.7 |
| Std. dev. | 90.3 | 22.3 | 10.6 | 5.8 |
| 5th percentile | 0.01 | 10.5 | 0.24 | 1.8 |
| 50th percentile | 0.27 | 16.7 | 1.95 | 3.5 |
| 95th percentile | 14.2 | 40.3 | 17.0 | 11.9 |
| *(-) Daily new COVID-19 cases (per $10^7$) [Lognormal]* | | | | |
| Mean | 1.29 | 11.8 | 1.11 | 1.8 |
| Std. dev. | 1.28 | 4.2 | 5.33 | 0.4 |
| 5th percentile | 0.25 | 2.7 | 0.10 | 0.3 |
| 50th percentile | 0.92 | 6.7 | 0.45 | 0.7 |
| 95th percentile | 3.53 | 7.8 | 3.70 | 2.7 |
| *Mask compulsory? [Normal]* | | | | |
| Mean | 1.45 | 20.2 | 0.35 | 1.8 |
| Std. dev. | 1.33 | 21.8 | 1.08 | 13.2 |
| 5th percentile | -0.74 | -6.9 | -1.43 | -6.1 |
| 50th percentile | 1.45 | 20.2 | 0.35 | 1.8 |
| 95th percentile | 3.64 | 26.4 | 2.13 | 9.2 |
| *Vaccine adoption (%) [Lognormal]* | | | | |
| Mean | 2.65 | 23.3 | 2.86 | 15.4 |
| Std. dev. | 7.42 | 13.0 | 16.1 | 13.4 |
| 5th percentile | 0.08 | 4.7 | 0.02 | 3.4 |
| 50th percentile | 0.90 | 8.6 | 0.51 | 5.5 |
| 95th percentile | 10.1 | 26.8 | 10.8 | 12.9 |

### 3.3 Sensitivity analysis

To illustrate the practical significance of the attributes and the extent of preference heterogeneity, we create predicted probability plots using the parameter estimates of specification 2 of MNL and MXL. The calculation of the predicted probabilities is based on the incremental logit model that capitalises on the linear-in-parameter utility specification (Chapter 5, Ben-Akiva and Lerman 2018). We assign the attributes of alternative 1 as the average of the difference in attributes of alternatives 1 and 2 across the sample, and attributes of alternative 2 are set to zero. To find the effect of a specific attribute on the propensity to travel by Underground with MNL, we only vary this attribute for alternative 1 and predict the probability to choose alternative 1 by plugging MNL estimates in Eq. 3. For MXL, we take 10,000 draws from the mixing distribution of utility parameters, compute the probability to choose alternative 1 for each draw, and summarize the distribution of the predicted probabilities. These probability plots show the change in propensity to use the Underground due to variation in a specific attribute.

Figure 2 shows the main effects of the attributes and overlays the MNL estimates on the MXL estimates. The mean predicted probabilities from the MXL model are closely aligned with those from the MNL model. Whereas mandatory face masks (on average) increase the probability of taking the Underground from 0.39 to 0.65, this probability increases from 0.50 to 0.75 when going from the no-vaccination to the full-vaccination scenario. Crowding density also plays a critical role as the predicated probability decreases from 0.49 to 0.12 due to a shift in crowding level from no crowding to technical capacity. The effects of travel time and new COVID-19 incidences are relatively smaller. The average probability of taking the Underground decreases from 0.40 to 0.28 due to an increase in travel time from 30 minutes to 90 minutes, and from 0.49 to 0.37 due to a shift from no COVID-19 cases to the highest COVID-19 case scenario.

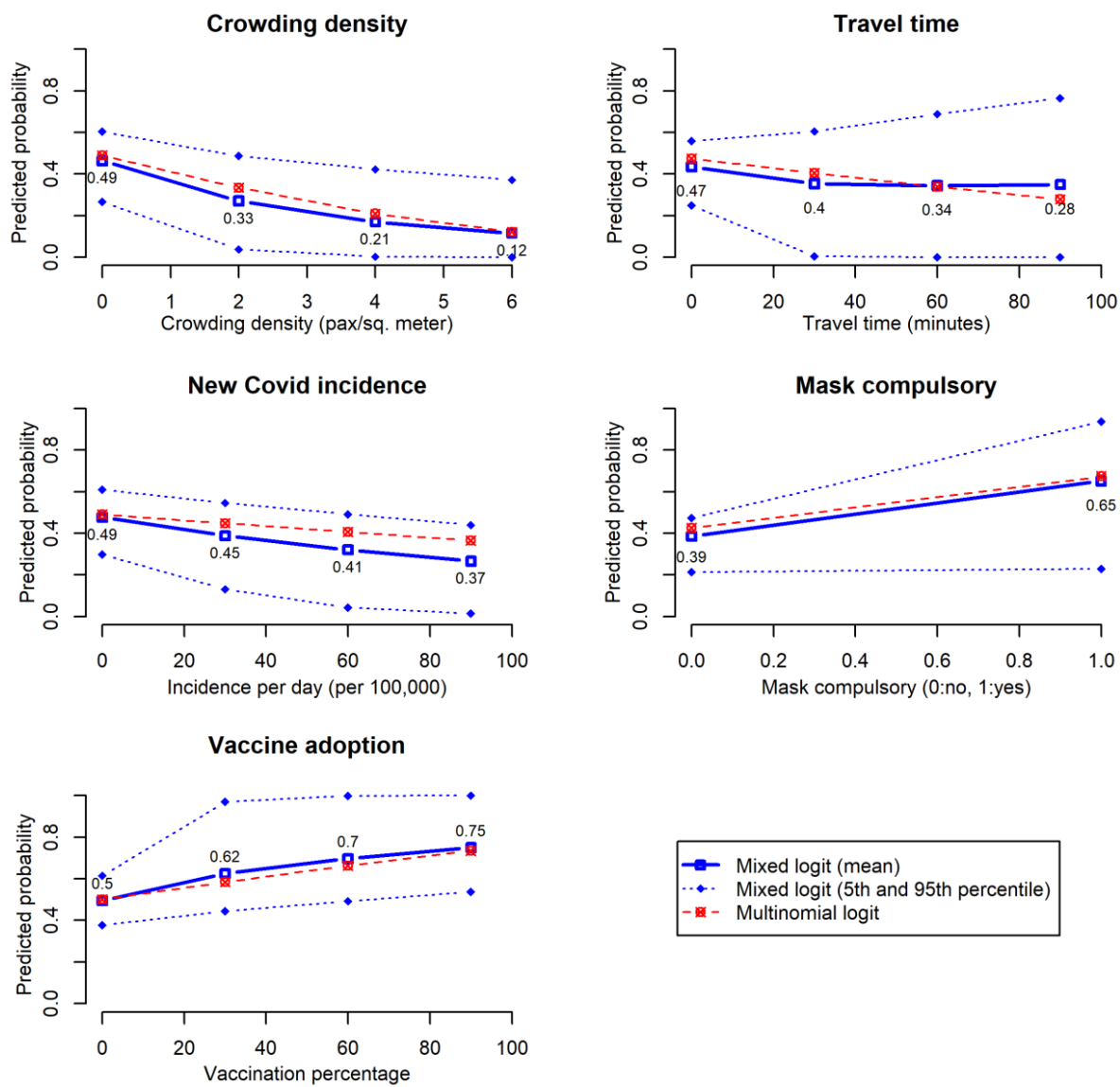

**Figure 2.** Main effects of the attributes on the probability of choosing the London Underground.

Figure 3 shows the effect of interactions between alternative-specific attributes on the predicted MXL probabilities. The average probability of choosing the Underground increases by 0.24 due to mandatory face masks for a 30-minute trip, but this increment reduces to 0.17 for a 90-minute trip (see Figure 3a). Preference heterogeneity for vaccination rates is prominent across crowding levels. The probability of choosing the uncrowded London Underground increases by 0.22 due to a shift from no-vaccination to the full-vaccination scenario, but this increment is 0.43 for the Underground operating at its technical capacity (see Figure 3b). The potential reasons behind such heterogeneity are discussed for the MNL results in Section 3.1, which also apply here.

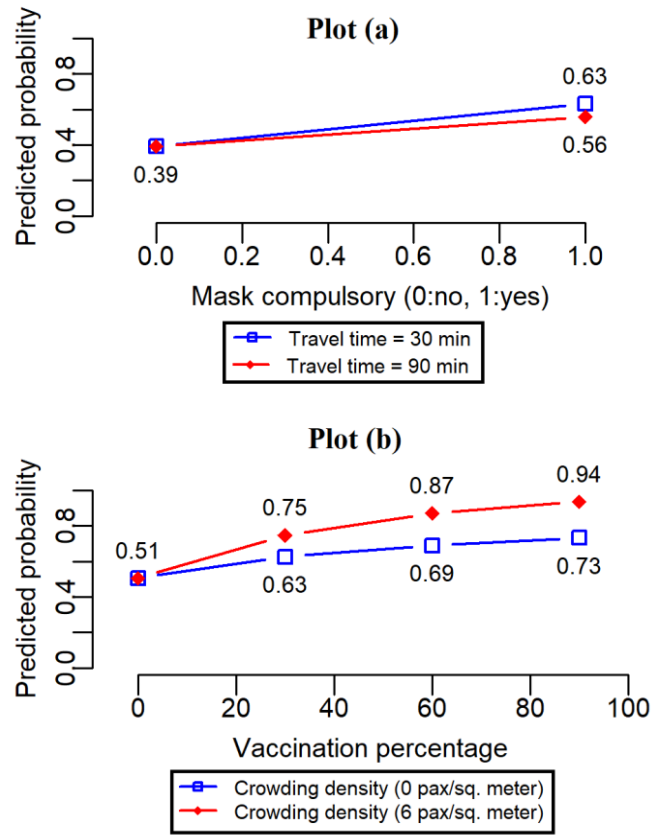

**Figure 3.** Heterogeneity in the effect of mandatory face masks and rates of vaccination on the probability of choosing the London Underground across different travel time values and crowding levels, respectively.

Figures 4, 5, and 6 present the heterogeneity in the mean attribute effects across individual-specific characteristics. The positive effect of mandatory face masks on the likelihood of taking the Underground is more pronounced among unemployed/retired respondents with age above 40 years relative to their counterparts (probability increment: 0.35 vs. 0.19, Figure 4a). Similarly, female respondents with monthly income below 10,000 pounds have a marginally higher positive impact of mandatory face masks than their counterparts (probability increment: 0.30 vs. 0.21, Figure 4b). The most noticeable heterogeneity in the effect of mandatory face masks is across respondents segmented based on their opinions about mask usage. Mandatory

face masks leads to an increase in the probability of choosing the Underground from 0.44 to 0.78 for mask supporters[2], but a decrease in probability to 0.3 for counterparts (see Figure 4c).

The positive effect of vaccination rates on the likelihood of using the London Underground is higher for respondents who identify as Asian or Asian-British. The increase in probability to use the Underground for this group of respondents due to a shift from no-vaccination to the full-vaccination scenario is 0.35, but the increment reduces to 0.23 for their counterparts (Figure 5a). As the UK moves towards a full-vaccination scenario, respondents with the opinion that masks help in controlling the spread of COVID-19 have a more striking increase in the likelihood to use the Underground than their counterparts (probability increment: 0.33 vs. 0.10, Figure 5b). A similar observation is made for pre-pandemic users with positive opinions[3] about vaccination and their counterparts (probability increment: 0.35 vs. 0.13, Figure 5c).

Figure 6a) shows that full-time employees who do not wear a mask in public spaces have a negligible negative effect of the daily new COVID-19 cases on their likelihood to use the Underground, but this effect is highly negative for their counterparts (probability decrement: 0.02 vs. 0.23). The likelihood to use the Underground by respondents who are older than 40 years, are employed, and hold at least a bachelor's degree is more negatively affected than their counterparts as the London Underground transitions from no-crowding to its technical capacity (probability decrement: 0.32 vs. 0.19, Figure 6b). Similarly, the propensity to use the Underground of respondents who have tested COVID-positive and are satisfied with the UK government's approach towards COVID-19 is less negatively affected due to the crowding density than their counterparts (probability decrement: 0.25 vs. 0.37, Figure 6c).

---

[2] The mask supporters agree that a mask helps in controlling COVID-19 spread and does not cause respiratory issues. They always wear a mask in public spaces.
[3] Respondents with a positive opinion about vaccines think that vaccines will be powerful in the long term, and they are worried about the side-effects of vaccines.

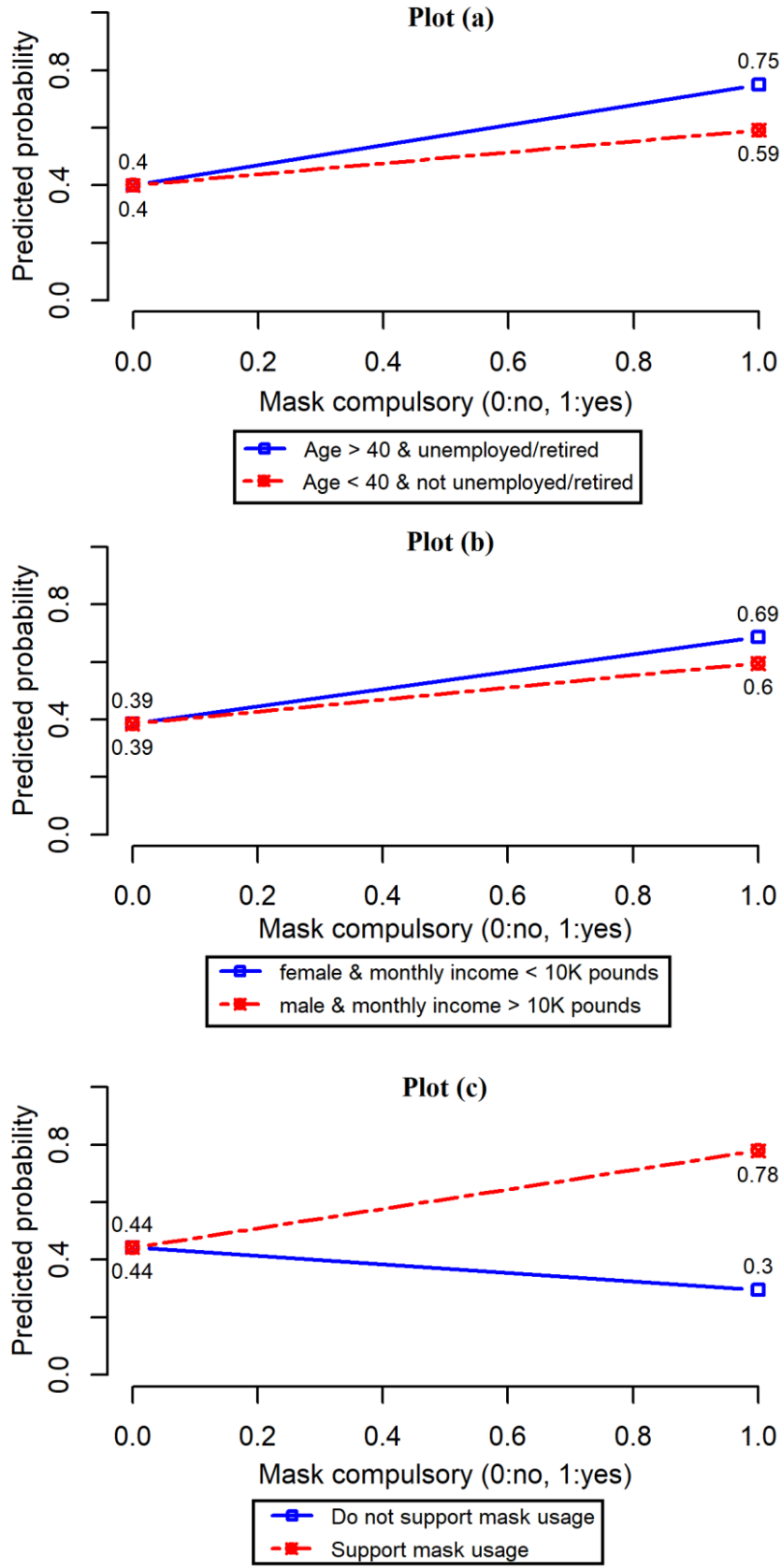

**Figure 4.** Heterogeneity in the effect of mandatory face masks on the probability of choosing the London Underground across groups with different demographics and opinions.

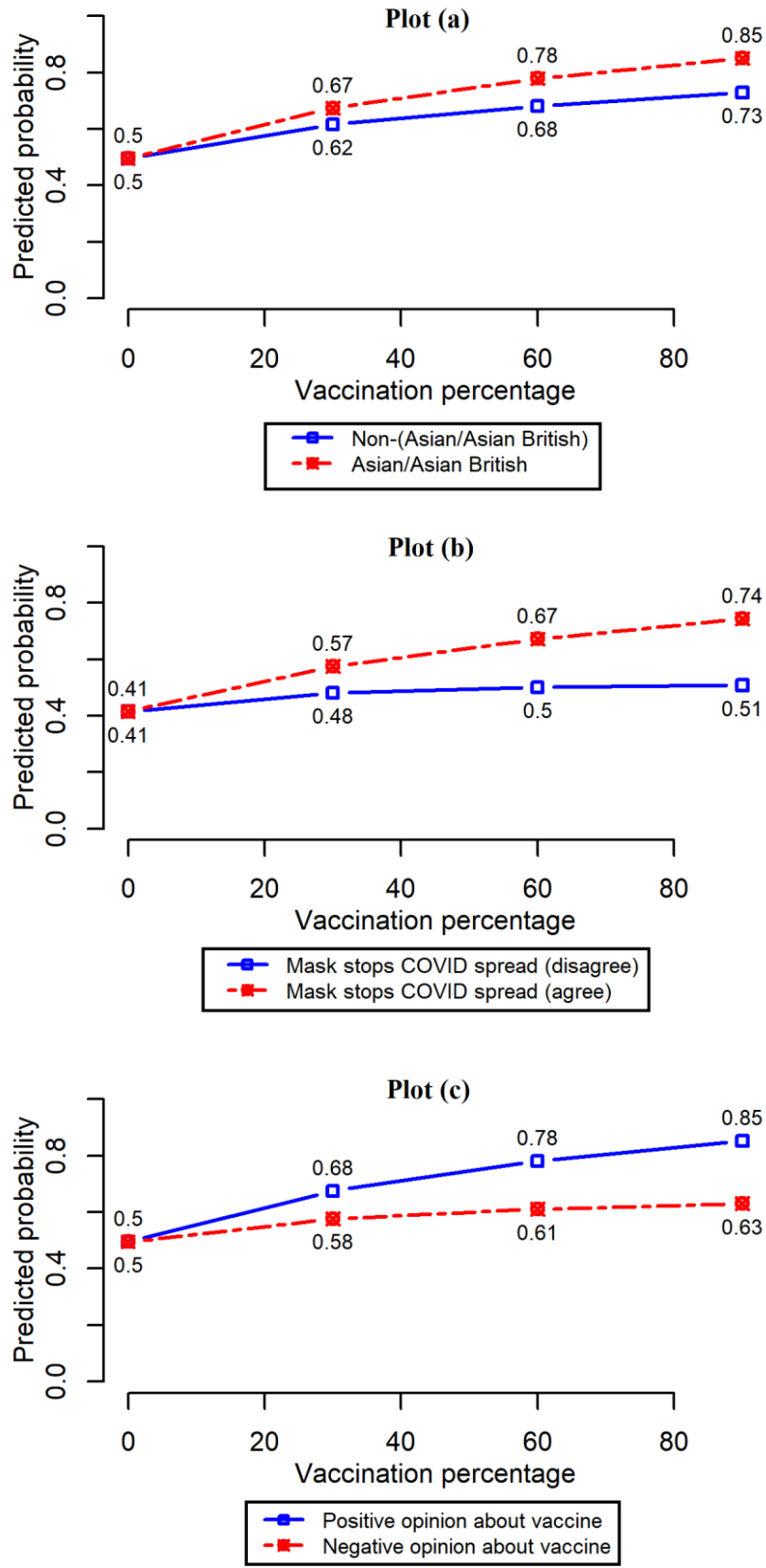

**Figure 5.** Heterogeneity in the effect of vaccination rates on the probability of choosing the London Underground across groups with different demographics and opinions.

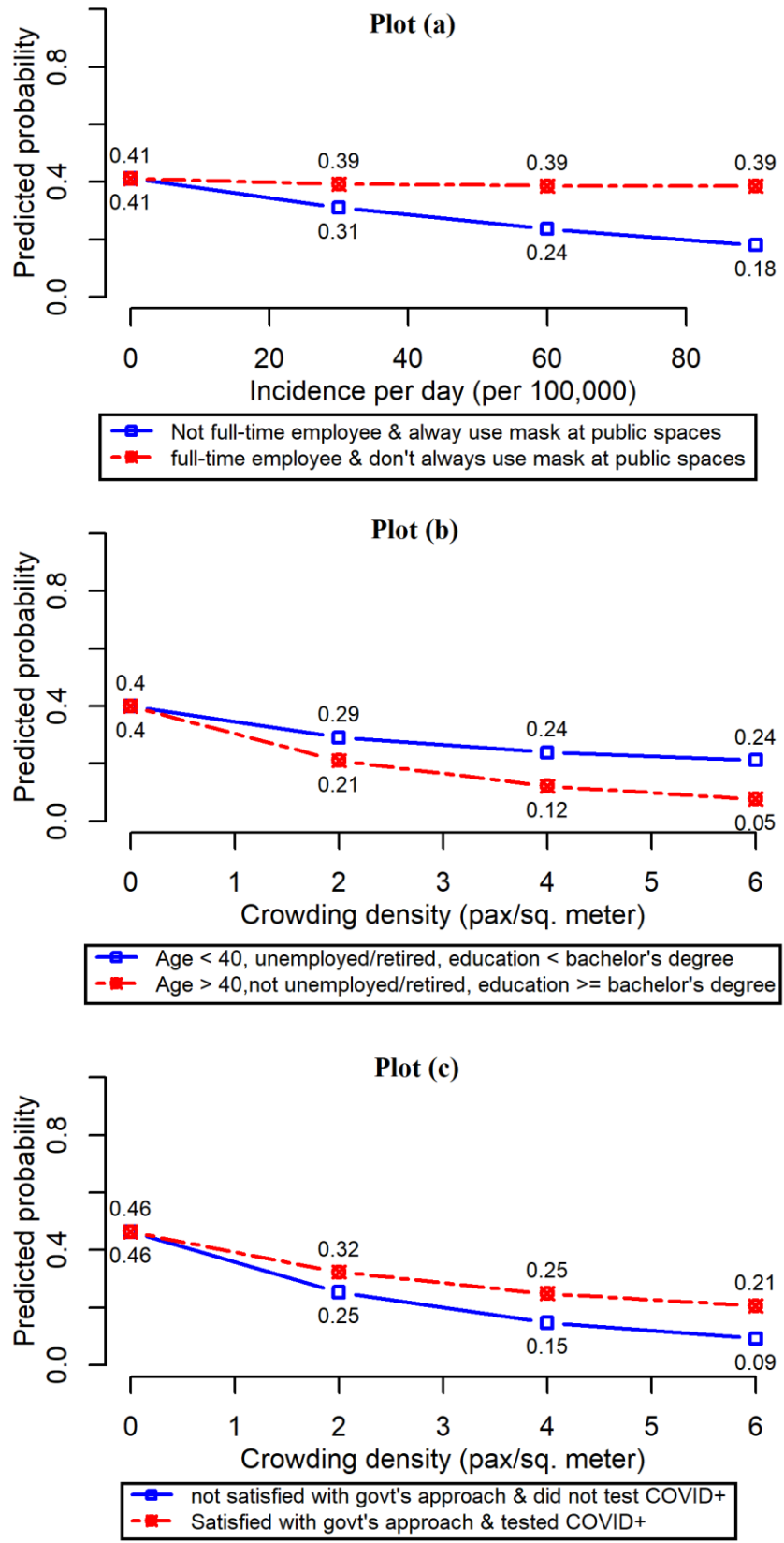

**Figure 6.** Heterogeneity in the effect of new COVID-19 incidences and crowding density on the probability of choosing the London Underground across groups with different demographics and opinions.

# 4 Conclusions

The COVID-19 pandemic has severely impacted the demand for public transport in many parts of the world. Notwithstanding the gradual easing of restrictions and increasing rates of vaccination, future demand for public transport remains uncertain. This study investigated the factors that influence travel demand for the London Underground during the pandemic. To that end, we conducted a stated choice experiment among pre-pandemic users of the London Underground and analysed the collected data using discrete choice models. Our study offers insights into the sensitivities of pre-pandemic users to travel by London Underground during the pandemic relative to travel attributes (crowding density, travel time), the epidemic situation (new confirmed COVID-19 cases), non-pharmaceutical interventions (whether masks are mandatory or not) and pharmaceutical interventions (vaccination rates).

Several implications for policy and practice can be derived from our analysis. First, since most travellers favour mandatory face masks, they should be made mandatory to encourage the use of public transport during the pandemic. Consistent with our findings, the World Health Organization also advises the mandatory face masks in public areas (WHO, June 2021). However, for longer trips, the wearing of face masks may cause discomfort to some travellers. Thus, if the goal is to encourage the use of public transport for longer trips, adequate incentives (e.g., free last-mile connectivity with bike-sharing systems) should be in place to compensate travellers for their discomfort. Second, as travellers are sensitive to crowding levels, travellers might adjust their plans in response to information about expected crowding levels. Therefore, advanced traveller information systems which provide updates on expected crowding levels may effectively reduce overcrowding and the risk of contagion on public transport. Third, our findings reveal that the perception of users about masks, vaccines, and the government's approach towards COVID-19 determines the extent to which countermeasures affect their propensity to use public transport. The effect of countermeasures also varies across socio-demographic characteristics. Thus, designing targeted advertising campaigns about the importance of masks, safety and efficacy of vaccines, and transparency in the government's approach is critical to recovering the demand for public transport.

There are multiple ways in which future research can build upon the work presented in this paper. First, the data for this study were collected between March and May 2021. Thus, the data only capture a snapshot of people's preferences and perceptions in a dynamically evolving environment. As respondents provided their answers under the impression of recent events, namely the gradual easing of lockdown restrictions and the ramp-up of vaccination campaigns, it would be instructive to repeat the analysis with data collected at a future point in time. Likewise, the study could be replicated in a different geographical context to evaluate the transferability of the results to other public transport systems. Second, at a more general level, our study underlines the importance of accounting for epidemic factors and non-pharmaceutical interventions in mathematical models of activity-travel behaviour. To plan for the "new normal" and possible future pandemics, the analysis presented in this paper may be extended to other components of travel behaviour (e.g., destination and detailed travel mode preferences). Finally, the sensitivities derived in this study may inform the development of epidemic control strategies and may be integrated into advanced person-centric epidemiological modelling and simulation frameworks.

## Acknowledgements

The data collection for this study is supported by Leverhulme Trust Early Career Fellowship (ECF-2020-246).

## CRediT Author Statement

*P Bansal:* Conceptualization, Methodology, Data Curation, Software, Writing - Original Draft
*R Kessels:* Conceptualization, Methodology, Software, Writing - Original Draft
*R Krueger:* Conceptualization, Methodology, Writing - Original Draft
*DJ Graham:* Writing - Review & Editing, Resources

# Appendix – Partial profile design of the discrete choice experiment (DCE)

The design of the DCE involved three blocks and eight choice situations per block. Each choice situation had two possible travel profiles of the London Underground. One of three blocks was randomly chosen and presented to the respondent. The design appears in Table A.1. The design accounted for the independent estimation of all main effects of the six attributes and all two-way interaction effects between them. The following restrictions were incorporated in the design: first, the two highest levels of daily new COVID-19 cases (70 and 90 per $10^5$) are incompatible with the three highest vaccination rates (50%, 65%, and 80%), and second, the two highest crowding levels (4 and 6) are not aligned with the three lowest vaccination rates (5%, 20%, and 35%) and the two highest levels of daily new COVID-19 cases (70 and 90 per $10^5$).

The choice situations contained partial profiles that were described by four varying attributes and two attributes with constant levels. For illustrative purposes, Table A.1 shows the varying attributes in grey. The constant attributes enable the estimation of the attribute interaction effects. In a block, each attribute is held constant in two or three choice sets and varied in five or six choice sets. The design was created using the partial profile design algorithm of Kessels et al. (2015) in the JMP Pro 16 software (SAS Institute Inc, Cary, NC, USA).

The design is Bayesian D-optimal, meaning that it incorporates all available knowledge about respondents' preferences in optimizing the D-criterion value to obtain the design that guarantees the most precise preference estimates. Defining priors was straightforward for the continuous attributes. That is, lower levels of crowding, in-vehicle (on-board) travel time, and daily new COVID-19 cases in the UK are generally preferred for travelling by the London Underground than higher levels for these attributes. The opposite is the case for the vaccinated population in the UK. We did not provide any prior preference regarding the standing and mask attributes as both preference directions on these attributes could be possible in COVID-19 times. Also, we allowed for quite some uncertainty or variability regarding all prior beliefs in the design optimization.

**Table A.1.** Partial profile design with three surveys (blocks).

| Survey | Choice situation | Crowding density | Standing in Underground? | Travel time ratio | New COVID-19 cases | Mask compulsory? | Vaccine adoption |
|---|---|---|---|---|---|---|---|
| 1 | 1 | 0 | No | 1.15 | 70 | No | 35% |
| 1 | 1 | 0 | No | 0.7 | 90 | Yes | 20% |
| 1 | 2 | 0 | Yes | 0.7 | 30 | Yes | 80% |
| 1 | 2 | 1 | Yes | 1 | 30 | No | 20% |
| 1 | 3 | 6 | Yes | 0.7 | 10 | No | 50% |
| 1 | 3 | 4 | No | 0.7 | 50 | Yes | 50% |
| 1 | 4 | 4 | No | 1.3 | 10 | Yes | 50% |
| 1 | 4 | 2 | Yes | 1.15 | 10 | Yes | 5% |
| 1 | 5 | 1 | Yes | 1.15 | 50 | No | 5% |
| 1 | 5 | 1 | No | 0.7 | 70 | Yes | 5% |
| 1 | 6 | 0 | No | 1.3 | 70 | No | 20% |
| 1 | 6 | 4 | Yes | 1.3 | 30 | No | 65% |
| 1 | 7 | 2 | Yes | 1.15 | 50 | No | 80% |
| 1 | 7 | 2 | No | 1.15 | 90 | Yes | 35% |
| 1 | 8 | 0 | Yes | 1.3 | 30 | No | 20% |
| 1 | 8 | 1 | Yes | 0.7 | 50 | Yes | 20% |
| 2 | 9 | 2 | Yes | 1 | 10 | Yes | 50% |
| 2 | 9 | 2 | No | 1.3 | 10 | No | 5% |
| 2 | 10 | 1 | No | 0.7 | 50 | Yes | 50% |
| 2 | 10 | 6 | No | 1 | 10 | Yes | 80% |
| 2 | 11 | 2 | Yes | 1.3 | 50 | No | 65% |
| 2 | 11 | 6 | No | 1.15 | 50 | Yes | 65% |
| 2 | 12 | 2 | Yes | 1.3 | 90 | No | 35% |
| 2 | 12 | 0 | Yes | 1.3 | 50 | Yes | 5% |
| 2 | 13 | 2 | No | 0.7 | 30 | Yes | 65% |
| 2 | 13 | 1 | Yes | 0.7 | 30 | No | 5% |
| 2 | 14 | 1 | No | 1.15 | 10 | Yes | 5% |
| 2 | 14 | 1 | Yes | 1.3 | 50 | Yes | 80% |
| 2 | 15 | 4 | Yes | 1.15 | 50 | No | 80% |
| 2 | 15 | 6 | No | 1.3 | 30 | No | 80% |
| 2 | 16 | 0 | Yes | 1.15 | 10 | Yes | 20% |
| 2 | 16 | 0 | Yes | 1 | 70 | No | 35% |
| 3 | 17 | 4 | Yes | 1.3 | 10 | No | 50% |
| 3 | 17 | 1 | No | 1.3 | 30 | Yes | 50% |
| 3 | 18 | 1 | No | 1 | 50 | No | 80% |
| 3 | 18 | 1 | Yes | 0.7 | 30 | No | 65% |
| 3 | 19 | 0 | Yes | 1 | 70 | No | 5% |
| 3 | 19 | 2 | No | 1.3 | 70 | Yes | 5% |
| 3 | 20 | 0 | No | 1 | 50 | Yes | 5% |
| 3 | 20 | 0 | Yes | 1.15 | 50 | No | 35% |
| 3 | 21 | 1 | No | 1.15 | 30 | No | 80% |
| 3 | 21 | 0 | No | 1.15 | 50 | Yes | 50% |
| 3 | 22 | 4 | Yes | 1.3 | 10 | Yes | 65% |
| 3 | 22 | 4 | No | 1.15 | 50 | No | 65% |
| 3 | 23 | 6 | No | 0.7 | 30 | No | 50% |
| 3 | 23 | 0 | Yes | 0.7 | 30 | Yes | 65% |
| 3 | 24 | 1 | Yes | 1.3 | 50 | Yes | 65% |
| 3 | 24 | 2 | Yes | 0.7 | 90 | Yes | 20% |